\documentclass[prb,floats,superscriptaddress]{revtex4}

\usepackage{graphicx}
\usepackage{epstopdf}
\pdfoutput=1
\usepackage{amssymb}
\usepackage{amsmath,bm}
\usepackage{psfrag}
\usepackage{epsfig}
\usepackage{float}

\newcommand{\angstrom}{\mbox{\normalfont\AA}}

\begin{document}
\title{Interplay of Quantum Size Effect, Anisotropy and Surface Stress Shapes the Instability of Thin Metal Films}

\author{Mikhail Khenner}
\affiliation{Department of Mathematics and Applied Physics Institute, Western Kentucky University, Bowling Green, KY 42101}

\begin{abstract}

Morphological instability of a planar surface ([111], [011], or [001]) of an ultra-thin metal film is studied in a parameter space formed
by three major effects (the quantum size effect, the surface energy anisotropy and the surface stress) that influence a film dewetting. The analysis is based on the extended Mullins equation, 
where the effects are cast as functions of the film thickness.
The formulation of the quantum size effect (Z. Zhang \textit{et al.}, PRL {\bf 80}, 5381 (1998)) 
includes the oscillation of the surface energy with thickness caused by electrons confinement. 
By systematically comparing the effects, their contributions into the overall stability (or instability) is highlighted.
\vspace{0.2cm}\\
\textit{Mathematics Subject Classification:}\ 74H10,\ 74H55,\ 35Q74.
\vspace{0.2cm}\\
\textit{Keywords:}\ Ultra-thin metal films, dewetting, PDE model, stability analysis.
\end{abstract}

\date{\today}
\maketitle


\section{Introduction}
\label{Intro}

Morphological instability of a surface of an ultra-thin solid film and film dewetting are important from the technology viewpoint, as well as from the 
viewpoint of basic physics. These phenomena were intensely studied in a semiconductor-on-semiconductor, metal-on-semiconductor, and metal-on-metal systems
\cite{Smith,Zhang,Hirayama,Ozer,Sanders,Han1}.
However, only for former systems were the mathematical models sufficiently developed and were able to help clarify the key mechanisms by which
the film becomes unstable and dewets; see, for instance, Refs. \cite{Chiu,GolovinPRB2004,LevinePRB2007,Korzec,Korzec1,KTL} and the references therein.

In this paper, building upon the just cited and closely related works, we develop a physico-mathematical model for the analysis of a morphology evolution of an ultra-thin metal films 
(thickness 5 to 30 monolayers (ML) (1.5-10 nm)) on a semiconductor or metal substrates. 
Our model accounts for three effects that are believed to play a major roles in a film instability and dewetting: the quantum size effect (QSE) \cite{Zhang}, 
the surface energy anisotropy, and the surface stress.

The advantage of a mathematical model is that it allows to separate the physical effects and thus clarify 
the contribution of each effect in the overall picture. This is seldom possible in experiment. For instance, the role in the initial surface instability of the surface energy that oscillates in the film thickness
\cite{Zhang,Ozer,Han,Han1} was not studied in detail. This paper partially fills this gap. We will refer to the surface energy oscillation as the QSE oscillation. (Such oscillation emerges primarily as a result of discreteness of the electronic energy bands due to the confinement at the film/vacuum
and the film/substrate interfaces. If the Fermi wavelength $\lambda_F$ has a commensurate relation with the monolayer height $\ell_0$ then, as the film thickness grows, new bands emerge under Fermi
level periodically, producing oscillations in the interface energy with period around $\lambda_F/2$. Another source of the surface energy oscillation is the Friedel oscillations 
in the electron density, which result in an oscillatory mean-field potential for electrons. The periodicity of the Friedel oscillation is also around $\lambda_F/2$.)

To the best of our knowledge, the 
existing stability results \cite{Zhang,Hirayama} for the ultra-thin metal films are based on the analysis of the second derivative (with respect to the film thickness) of the 
free-energy function. Such analysis is not without drawbacks, e.g., the perturbation wavelength and its 
characteristic development time scale are not easy to predict. The partial differential equation (PDE)-based model allows not only the calculation of these quantities (that are important in self-assembly), but also a
computation of the evolution of the nano-pits \cite{Srol,K1,K2} and the retraction of the film edge 
\cite{Thompson,Wong1,Wong2}.

The paper proceeds as follows. In Sec. \ref{Formulation} we first present the general form of the governing PDE; this PDE is the classical Mullins' surface diffusion equation, 
which includes the three above-mentioned effects. \footnote{Herring-Mullins theory has been used to study the instabilities and morphological evolution of a comparatively thick films 
($>$ 100 nm). However, the theory does not have a limitation on a thickness \textit{per se}. The base assumption in the theory is that there is a mobile adsorbate on the crystal surface, i.e. the top layer of the crystal is ``fluid", and the geometrical quantities such as the curvature can be defined for the surface. The top layer is usually considered to be one to three atoms deep. Thus the theory is applicable to thinner films. It must be noted here that the dewetting of the ultra-thin metallic films  is apparently driven by surface diffusion, as the films develop a faceted pinholes that extend from the surface down to the substrate, and the pinholes are surrounded at their surface perimeter by the characteristic high rims 
formed out of the film material that diffused from the pinhole bottom \cite{Sanders}.} Then we focus on each effect separately and arrive to their mathematical formulations, that are 
next inserted into the governing PDE. In Sec. \ref{DerivePDE} we employ a perturbation method and derive the final form of the PDE. In this paper we present only those
terms of the PDE that contribute to the linearized form. The linear stability analysis (LSA) is performed in Sec. \ref{LSA}; here we again focus separately on each effect
and then take a brief look at the situation where all three effects are present. Section \ref{Concl} contains the conclusions.

\section{Problem formulation}
\label{Formulation}

The starting point for the model is Mullins' surface diffusion equation: 
\begin{equation}
h_t=\sqrt{1+h_x^2+h_y^2}\ \mathcal{D}\nabla_s^2 \mu,
\label{MullinsEq}
\end{equation}
where $h(x,y)$ is the height of the surface of  the thin film above the substrate, $x,\;y$ the coordinates in the 
substrate plane, 
$\nabla_s^2$ the surface Laplacian, $\mathcal{D}=\Omega^2 D \nu/k T$ the diffusion constant ($\Omega$ is the adatom volume, $D$ the adatom diffusivity, 
$\nu$ the surface density of the adatoms, $k T$ the Boltzmann's factor),
and $\mu=\frac{\delta}{\delta h}\int \gamma\; dS$ the chemical potential;
here $\gamma$ is the surface energy, $\frac{\delta}{\delta h}$ the functional derivative and the integral is over the film surface.
Also let $z$ be the coordinate normal to the substrate plane.

Due to crystalline anisotropy, the surface energy depends on the surface orientation ${\bf n}$, where 
\begin{equation}
{\bf n}=(n_1,n_2,n_3)=\left(\frac{-h_x}{\sqrt{1+h_x^2+h_y^2}},\frac{-h_y}{\sqrt{1+h_x^2+h_y^2}},\frac{1}{\sqrt{1+h_x^2+h_y^2}}\right)
\label{n}
\end{equation}
is the unit normal to the surface. QSE and the surface stress (discussed later in this section) introduce a dependence of the surface energy on $h$, i.e. $\gamma = \gamma({\bf n},h)$.
Then the calculation of the functional derivative leads to a four-term expression \cite{Korzec} (see also Refs. \cite{Korzec1,GolovinPRB2004,LevinePRB2007,KTL})
\begin{equation}
\mu = \mu_\kappa+\mu_{wet}+\mu_{anis}+\mu_{h.o.t.},
\label{mu}
\end{equation}
with
\begin{eqnarray}
\mu_\kappa &=& \gamma \kappa,\quad \kappa = n_3^3\left[-h_{xx}\left(1+h_y^2\right)-h_{yy}\left(1+h_x^2\right)+2h_xh_yh_{xy}\right], \label{mu_kappa}\\
\mu_{wet} &=& n_3 \partial_h \gamma, \label{mu_wet}\\
\mu_{anis} &=& -2n_3\left[\left(h_x h_{xx}+h_yh_{xy}\right)\partial_{h_x}\gamma+\left(h_y h_{yy}+h_xh_{xy}\right)\partial_{h_y}\gamma\right]-\frac{\partial_x\partial_{h_x}\gamma+
\partial_y\partial_{h_y}\gamma}{n_3}, \label{mu_anis}\\
\mu_{h.o.t.} &=& \partial_{xx}\partial_{h_{xx}}\left(\frac{\gamma}{n_3}\right)+\partial_{xx}\partial_{h_{yy}}\left(\frac{\gamma}{n_3}\right)+
\partial_{yy}\partial_{h_{xx}}\left(\frac{\gamma}{n_3}\right)+\partial_{yy}\partial_{h_{yy}}\left(\frac{\gamma}{n_3}\right),
\label{mu_hot}
\end{eqnarray}
where $\kappa$ is the curvature, $n_3$ is seen in Eq. (\ref{n}) and ``wet", ``anis", and ``h.o.t." are the short-hand notations for wetting, anisotropy and higher order terms, respectively.

The expression for the surface Laplacian in Eq. (\ref{MullinsEq}) reads:
\begin{equation}
\nabla_s^2 = n_3^2\left[\left(1+h_y^2\right)\partial_{xx}+\left(1+h_x^2\right)\partial_{yy}-2h_xh_y\partial_x\partial_y+\kappa\frac{h_x\partial_x+h_y\partial_y}{n_3}\right],
\label{surfLaplace}
\end{equation}
and the surface energy in Eqs. (\ref{mu_kappa})-(\ref{mu_hot}) is the sum, 
\begin{equation}
\gamma = \gamma_0 + 
\gamma^{(QSE)} + \gamma^{(SS)} +\gamma^{(Anis)},
\label{totalEnergy}
\end{equation}
where $\gamma_0$ is the nominal (constant) surface energy, and the contributions labeled ``QSE", ``SS" and ``Anis" are due to quantum size effect, the surface stress and the anisotropy, respectively.
In sections \ref{QSE_sect}, \ref{SSen} and \ref{Ans}
we introduce and discuss these 
contributions.
We also remark that this paper is a general
theoretical study. It is not our goal to study any particular material system. 
Therefore the parameter values that we use are realistic (feasible) at best. 

\subsection{QSE contribution to surface energy}

\label{QSE_sect}

The non-interacting electron gas model \cite{OPL,Han,Hirayama} suggests the following form of $\gamma^{(QSE)}$
for many metal-on-semiconductor \cite{Smith,Zhang,Hirayama,Ozer} (Ag/GaAs, Ag/Si, Pb/Si, and Pb/Ge) and metal-on-metal systems \cite{Han1} (Pb/Cu, Ag/Fe, Ag/NiAl, and Fe/CuAu):
\begin{equation}
\gamma^{(QSE)}(h) = \gamma_0^{(QSE)} + \frac{g_0 s^2}{(h+s)^2}\cos{\eta h}-\frac{g_1 s}{h+s},\quad g_0, g_1 > 0,
\label{gammaQSE}
\end{equation}
where the first (constant) term and the second term are due to quantum confinement of the electrons in a thin film, and the third term is due to the
electrons spilling out to the film/substrate interface (the charge spilling effect). 
Eq. (\ref{gammaQSE}) results in the limit of thick film of a \emph{quantum} model, meaning that $N\gg 1$, where $N$ is the number of energy bands occupied by the electrons 
in the $z$-direction. $\gamma_0^{(QSE)} = k_F^2 E_F/(80 \pi) \sim 400-1000$ erg$/$cm$^2$ and $\eta = 2k_F=4\pi/\lambda_F\sim 20$ nm$^{-1}$ for Ag, Au, Mg, Al and Pb. 
(Here $E_F$ is the energy at the bulk Fermi level.)
The energies $g_0$ and $g_1$ also are proportional to $k_F^2 E_F$ \cite{OPL}. 
A small wetting length $s$ ($\sim 4 \angstrom=0.4 \mbox{nm}\sim 1.5-2$ ML \cite{Chiu}) prevents divergence as $h\rightarrow 0$. 
The continuous form (\ref{gammaQSE}) is the result of the interpolation of the set of data points produced by the quantum model \cite{OPL,Han}, whose $z$-coordinates are the 
integer multiples of $\ell_0$ (the monolayer height); see for instance Eq. (17) in Ref. \cite{Han} and Figure 5(a) in that paper.
With the choice of $s$ and $\gamma_0$ for the height and energy scales, the dimensionless form of $\gamma^{(QSE)}(h)$ reads:
\begin{equation}
\gamma^{(QSE)}(H) = R_{\gamma_0}+\frac{G_0}{(1+H)^2}\cos{\rho H}-\frac{G_1}{1+H},
\label{gammaQSE2}
\end{equation}
where $R_{\gamma_0}=\gamma_0^{(QSE)}/\gamma_0,\; G_0=g_0/\gamma_0,\; G_1=g_1/\gamma_0$ and $\rho=s\eta$. Notice that $0 < R_{\gamma_0},\ G_0,\ G_1 < 1$, and $\rho \sim 10$.

%
\begin{figure}[h]
\centering
\includegraphics[width=3.0in]{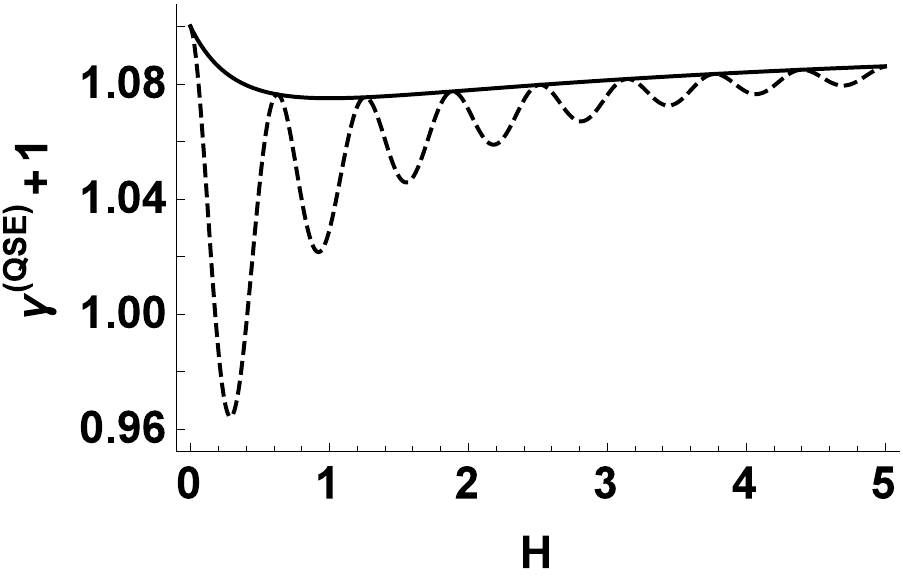}
\caption{Graphs of $1+\gamma^{(QSE)}(H)$, which is the nominal surface energy (the dimensionless value one) offset by $\gamma^{(QSE)}(H)$; $R_{\gamma_0}=G_0=G_1=0.1$. Solid line: $\rho=0$ (the QSE oscillation is absent). Dashed line: $\rho=10$.
}
\label{FigGammaQSE}
\end{figure}

From the viewpoint of the \emph{thermodynamic} stability analysis a planar surface of a film is stable, if $\frac{\partial^2}{\partial H^2}(1+\gamma^{(QSE)}(H)) \ge 0$.
In the absence of the QSE oscillation ($\rho=0$ in Eq. (\ref{gammaQSE2}))  this yields the condition
\begin{equation}
G_1 \le \frac{3G_0}{1+H_0}\; \Longleftrightarrow\; H_0 \le 3G_0/G_1-1 = H_{0c}.
\label{StabCondThermodyn}
\end{equation}  
Thus a deposited film with a planar surface is unstable, if its thickness $H_0$ exceeds the critical value $H_{0c}$; 
clearly, $H_{0c}$ increases linearly with the ratio $G_0/G_1$.
Notice that \emph{all} films are unstable if $G_0\le G_1/3$, since in this case 
$H_{0c}\le 0$. For the parameters
values in Figure \ref{FigGammaQSE} $H_{0c}=2$, and at this thickness the curvature of the energy curve (the solid line in Figure \ref{FigGammaQSE})
switches from a positive value to a negative one (the inflection point). 
Also, Hirayama \cite{Hirayama} points out that the minimum of the energy curve has to occur
at $H>0$. This results in the following restriction on $G_0$ and $G_1$: $G_1 < 2G_0$.

When the QSE oscillation is present, the thermodynamic criterion of stability can be only solved numerically for $H_{0c}$. 
A far better and more versatile method is a full linear stability analysis, which we perform in section \ref{LSA_QSE}. 

\subsection{Surface stress contribution to surface energy}
\label{SSen}

We assume that the lattice mismatch between the film and the substrate (and more generally, 
the interaction energy of a film and substrate atoms)  has a negligible effect on the
stress at the surface; in other words,
a thin metal film on a substrate is considered free-standing for the purpose of calculating the stress at the surface. The only source of stress is thus the 
intrinsic surface stress.\footnote{ It was argued recently \cite{Liu} that in addition to this conventional surface stress of a macroscopic thin film, the ultra-thin metal films also
are influenced by the quantum oscillations in the surface stress that are phase shifted with respect to the oscillations of the surface energy (the latter oscillations are termed 
the QSE oscillations in this paper).
The authors of Ref. \cite{Liu} propose that the coupling of the two oscillations may be responsible for the extension of the surface energy oscillations to thick 
films ($> 30$ ML).}
(It is well-known that 
the epitaxial
relation between the metal film and a substrate is often
unclear \cite{Sanders}, and in some systems (Ag/Fe(100)) the lattice mismatch is nearly absent \cite{Han1}. Hirayama \cite{Hirayama} points out that even when the misfit strain is large the electronic growth theory, in which the mismatch strain is not considered, 
is consistent with experiments on Ag/Si(111) system. He points out the possibility that the film-substrate bonding is weak, which allows the film to relax fully without introducing dislocations.)

\begin{figure}[h]
\centering
\includegraphics[width=3.0in]{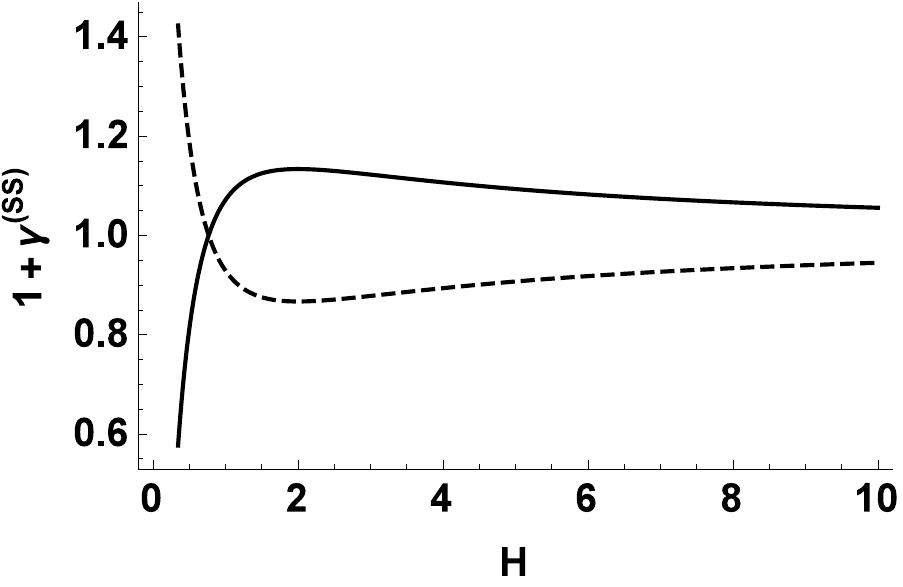}
\caption{Plot of $1+\gamma^{(SS)}(H)$. Parameters are: $\Gamma P=9.6\times 10^{-8}$ erg 
(corresponding to $P=2500$ atoms), $M = 31.2\times 10^{-12}$ erg/atom \cite{HW}, $\epsilon_*= 0.01,\; \ell = 10^{-5}$ cm =100 nm, $\gamma_0=1.8\times 10^3$ erg/cm$^2$
(corresponding to Cu[100] surface \cite{Tyson}). Solid line: $c_s=-1$ (compressive stress); dashed line: $c_s=1$ (tensile stress).
}
\label{FigGammaSS}
\end{figure}

In the 2009 paper, Hamilton \& Wolfer \cite{HW} re-visited the Gurtin-Murdoch model of surface elasticity \cite{GM} and derived the analytical relationship between the strain and 
the thickness of a free-standing metal nano-sheets.
This relationship makes possible the derivation of a surface stress energy. 
The starting point is the following expression from Sec. 5 of Ref. \cite{HW}, which holds in the case of 
equi-biaxial strain:
\begin{equation}
\gamma^{(SS)}= \left(\Gamma_{11}+\Gamma_{12}\right)\left(2\epsilon_*+\epsilon\right)\epsilon.
\label{gammaSS}
\end{equation}
Here $\Gamma_{11},\;\Gamma_{12}$ are the surface elastic moduli (with the units erg/atom), $\epsilon$ the strain (dimensionless) and 
$\epsilon_*$ the characteristic strain parameter, also dimensionless. 
Equation (\ref{gammaSS}) is thus the surface energy per atom.
Substitution in Eq. (\ref{gammaSS}) of $\epsilon$ from Eq. (4) of Ref. \cite{HW}, followed by the conversion to the energy per unit area gives
\begin{equation}
\gamma^{(SS)}(h)= \frac{-4 \epsilon_*^2 \left(\Gamma_{11}+\Gamma_{12}\right)^2 \left[\Gamma_{11} + 
   \Gamma_{12} + (h/\ell_0 - 2) \left(M_{11} + M_{12}\right)\right] P}{\left[2 \Gamma_{11} + 
   2 \Gamma_{12} + (h/\ell_0 - 2) \left(M_{11} + M_{12}\right)\right]^2 \ell^2},
\label{gammaSS1}
\end{equation}
where $M_{11},\;M_{12}$ are the bulk elastic moduli, $P$ the total number of atoms in the film, and $\ell$ is the (large) lateral
dimension of the film (the same along the $x$ and $y$ axes). Introducing 
the height scale replaces $h/\ell_0$ by $s H/\ell_0$. Using $\gamma_0$ again for the energy unit, and taking $s=\ell_0$, without a significant loss of generality we obtain the following dimensionless form:
\begin{equation}
\gamma^{(SS)}(H)=\frac{-F \Gamma \left[\Gamma + M (H-2)\right]}{\left[2 \Gamma + M (H-2)\right]^2},
\label{gammaSS2}
\end{equation}
where $\Gamma=\Gamma_{11}+\Gamma_{12}$, $M=M_{11}+M_{12}$ and $F=4 c_s \Gamma P \epsilon_*^2/(\gamma_0 \ell^2)$. 
(Notice that $F$ is dimensionless, and because $\Gamma$ and $M$ have a unit of energy per atom, this unit cancels from Eq. (\ref{gammaSS2}).)
$c_s=\pm 1$ has been introduced into $F$ to account
for tensile ($c_s=1$) or compressive ($c_s=-1$) stress. Figure \ref{FigGammaSS} shows that according to the thermodynamic stability criterion, thick films 
($H>3$ for the chosen set of parameters) 
subjected to a compressive surface stress are stable, while such films subjected to a tensile
surface stress are unstable.  Thin films $(H\le 3)$ are unstable under compressive stress and stable under tensile stress. \cite{McCarty} 

\subsection{Anisotropy contribution to surface energy}
\label{Ans}

To introduce anisotropy we follow the established approach \cite{McFadden}.  In the Cartesian coordinate system $(x',y',z')$ aligned with the crystal axes 
(the crystalline system) the 
cubic-symmetric surface energy of a surface element with a normal direction ${\bf n'}=\left(n_1',n_2',n_3'\right)$ in that system, is:
\begin{equation}
\gamma'\left(n_1',n_2',n_3'\right) = \gamma_0\epsilon_\gamma\left(n_1^{'4}+n_2^{'4}+n_3^{'4}\right),
\label{gammaAnis1}
\end{equation}
where $\epsilon_\gamma$ is the anisotropy strength. The unit normals in the crystalline and laboratory systems are related by a rotation,
\begin{equation}
\begin{pmatrix}
  n_1' \\
  n_2' \\
  n_3' 
\end{pmatrix}
=
\begin{pmatrix}
  u_{1,1} & u_{1,2} & u_{1,3} \\
  u_{2,1} & u_{2,2} & u_{2,3} \\
  u_{3,1} & u_{3,2} & u_{3,3}
\end{pmatrix}
\begin{pmatrix}
  n_1 \\
  n_2 \\
  n_3 
\end{pmatrix},
\label{RelateNormals}
\end{equation}
where for [001] orientation of the $z$-axis with respect to primed variables: $u_{1,1}=u_{2,2}=u_{3,3}=1$ and other elements of the rotation matrix are zeros;
for [011] orientation: $u_{1,1}=1,\; u_{2,2}=u_{2,3}=u_{3,3}=1/\sqrt{2},\; u_{3,2}=-1/\sqrt{2}$, other elements are zeros;
for [111] orientation: $u_{1,1}=\sqrt{2/3},\; u_{2,2}=1/\sqrt{2},\; u_{1,3}=u_{2,3}=u_{3,3}=1/\sqrt{3},\; u_{2,1}=u_{3,1}=-1/\sqrt{6},\; u_{3,2}=-1/\sqrt{2},\; u_{1,2}=0\;$ \cite{McFadden}.

The substitution of $n_1,\;n_2,\;n_3$ from Eq. (\ref{n}) into Eq. (\ref{RelateNormals}), followed by the substitution of $n_1',\;n_2',\;n_3'$ into Eq. (\ref{gammaAnis1}) 
and expansion in the powers of $h_x,\; h_y$, and then omitting the contributions that are proportional to
the powers higher than four (the small-slope approximation) gives

\begin{equation}
\gamma^{(Anis)}=\gamma_0 \epsilon_\gamma\left(a_{00} + a_{20} h_x^2 + a_{30} h_x^3 + a_{40} h_x^4 + a_{02} h_y^2 + 
      a_{04} h_y^4 + a_{12} h_x h_y^2  + a_{22} h_x^2 h_y^2
      \right), \label{anizcoeffs}
\end{equation}
where the coefficients $a_{ij}$ are shown in the Table 1. 
\begin{center}
    \begin{tabular}{| c | c | c | c | c | c | c | c | c | c |}
    \hline
    Orientation & $a_{00}$ & $a_{20}$ & $a_{30}$ & $a_{40}$ & $a_{02}$ & $a_{04}$ & $a_{12}$  & $a_{22}$  \\ \hline
    [001] & 1 & -2 & 0 & 4 & -2 & 4 & 0  & 6  \\ \hline
    [011] & $1/2$ & -1 & 0 & $5/2$ & 2 & -4 & 0 &  -3   \\ \hline
    [111] & $1/3$ & $4/3$ & $-2\sqrt{2}/3$ & $-5/2$ & $4/3$ & $-5/2$ & $2\sqrt{2}$  & -5   \\
    \hline
    \end{tabular}\vspace{0.5cm}\\
Table 1. Coefficients in Eq. (\ref{anizcoeffs}) for three typical orientations of the crystal surface.
\end{center}

Next, we introduce a small parameter $\alpha=s/\ell$. Thus $h=sH=\alpha \ell H$. Augmented by the following scaling of the in-plane coordinates: $(x,y)=\ell (X,Y)$, this is termed \emph{a thin film scaling}. 
After applying the thin film scaling 
and the energy scale $\gamma_0$, the dimensionless anisotropic surface energy reads
\begin{eqnarray}
\gamma^{(Anis)}\left(H_X, H_Y, ...\right) &=& \epsilon_\gamma\left[a_{00} +  \alpha^2\left(a_{20} H_X^2 + \alpha a_{30}  H_X^3 + \alpha^2 a_{40} H_X^4 + a_{02} H_Y^2 + 
      \alpha a_{12} H_X H_Y^2 \right.\right. \nonumber \\ 
&+& \left.\left. \alpha^2 a_{22} H_X^2 H_Y^2 + 
      \alpha^2 a_{04} H_Y^4\right)\right] + \frac{\Delta}{2}\bar \kappa^2.
\label{gammaAnis2}
\end{eqnarray}
The right-hand side of Eq. (\ref{gammaAnis2}) has been augmented by the term $(\Delta/2)\bar \kappa^2$, where  
\begin{equation}
\bar \kappa = \frac{-\alpha H_{XX} \left(1 + \alpha^2 H_Y^2 \right)- \alpha H_{YY} \left(1 + \alpha^2 H_X^2 \right)+2 \alpha^3 H_X H_Y H_{XY}}{\left(1 + \alpha^2 H_X^2  + \alpha^2 H_Y^2 \right)^{3/2}}
\label{curv_alpha}
\end{equation}
is the dimensionless curvature and $\Delta= \delta/(\gamma_0 \ell^2)$ the dimensionless parameter. Here $\delta$ is the corner energy.
As argued in Ref. \cite{GDN}, this curvature-squared ``regularization" (first noted by C. Herring \cite{Herring} and proposed by Gurtin \textit{et al.} \cite{CGP} in the context of 
a PDE-based model of a surface morphological evolution) can be traced to the interaction of steps in a surface region where two facets meet at a corner.

\section{Evolution equation for the film thickness}

\label{DerivePDE}

To derive the evolution PDE for the film thickness, we first substitute Eqs. (\ref{gammaQSE2}), (\ref{gammaSS2}) and (\ref{gammaAnis2}) 
into Eq. (\ref{totalEnergy}).
Then the resultant total dimensionless surface energy $\gamma$ is substituted into the system (\ref{MullinsEq}) - (\ref{surfLaplace}). We complete adimensionalization there using 
the thin film scaling augmented by the time scaling $t=(\ell^2/D)T$.
At this stage the dimensionless evolution PDE takes the form \cite{Korzec}:
\begin{equation}
H_T = \frac{B}{\alpha}\sqrt{1+\alpha^2\left(H_X^2+H_Y^2\right)}\left[\bar \nabla_s(\alpha,\ldots)^2\left\{\bar \mu_{\kappa}(\alpha,\ldots)+\bar \mu_{wet}(\alpha,\ldots)+\bar \mu_{anis}(\alpha,\ldots)+\bar \mu_{h.o.t.}(\alpha,\ldots)\right\}\right],
\label{H_T1}
\end{equation}
where $B=\Omega^2 \nu \gamma_0/\ell^2 k T$, the overbar denotes the dimensionless quantities, and the ellipsis stands for $H$, its partial derivatives, and the dimensionless parameters $F,\ R_{\gamma_0},\ G_0,\ G_1,\ \Gamma/M,\ \rho,\ a_{00}, \ldots, a_{22},\ \Delta$. 
To leading order in $\alpha$, $(B/\alpha)\sqrt{1+\alpha^2\left(H_X^2+H_Y^2\right)}\approx B/\alpha$, and 
we set $B/\alpha = 1$ here (thus the perturbation growth rate that we calculate and compute below is modulo $B/\alpha$).

Next, we expand $\bar \nabla_s^2 \bar \mu_{\kappa}$, $\bar \nabla_s^2 \bar \mu_{wet}$, $\bar \nabla_s^2 \bar \mu_{anis}$,
and $\bar \nabla_s^2 \bar \mu_{h.o.t.}$ in the powers of $\alpha$, retaining only the dominant order coefficients of all expansions. This finally yields a cumbersome highly nonlinear PDE. 
Since the goal of this paper is the linear stability analysis, we omit those terms in the PDE that are proportional to any product of the spatial derivatives of $H$.

Now the equation takes the form
\begin{equation} 
H_T = \sum_{i=1}^{6} f_i,
\label{PDEcompact}
\end{equation}
where the terms $f_i$ are as follows.
\begin{equation}
f_1=-H_{XXXX} - 2 H_{XXYY} - H_{YYYY};
\label{f_1}
\end{equation}
this is $\gamma_0$ contribution (see Eq. (\ref{totalEnergy})), which originates in $\bar \nabla_s^2 \bar \mu_{\kappa}$ (the traditional Mullins' terms).
\begin{eqnarray}
f_2 &=&  \frac{F \Gamma (\Gamma+M(H-2))}{(2 \Gamma + M(H-2))^2} \left(H_{XXXX} + 2 H_{XXYY} + 
    H_{YYYY}\right) \label{StressTerms} \\ 
&+& \frac{2F \Gamma M^2 (\Gamma - M(H-2))}{(2\Gamma + M(H-2))^4}  \left(H_{XX} + H_{YY}\right) \nonumber \\ 
&\equiv& q_{11}(H;\; F, \Gamma, M) \left(H_{XXXX} + 2 H_{XXYY} + H_{YYYY}\right) + q_{12}(H;\; F, \Gamma, M) \left(H_{XX} + H_{YY}\right); \nonumber
\end{eqnarray}
this is $\gamma^{(SS)}$ contribution, where the fourth (second) derivatives originate in $\bar \nabla_s^2 \bar \mu_{\kappa}$ $(\bar \nabla_s^2 \bar \mu_{wet})$.
\begin{eqnarray}
f_3 &=&  -\left(R_{\gamma_0}+\frac{G_0}{(1+H)^2}\cos{\rho H}-\frac{G_1}{1+H}\right)\left(H_{XXXX} + 2 H_{XXYY} + H_{YYYY}\right) \label{QSEterms}\\
&\equiv& - q_{21}\left(H;\; R_{\gamma_0},G_0,G_1,\rho\right)\left(H_{XXXX} + 2 H_{XXYY} + H_{YYYY}\right); \nonumber
\end{eqnarray}
this is $\gamma^{(QSE)}$ contribution, which originates in $\bar \nabla_s^2 \bar \mu_{\kappa}$. 
\begin{eqnarray}
f_4 &=&
\left[\frac{-2G_1}{(1 + H)^3}+\frac{G_0}{(1 + H)^2}\left(\frac{6}{(1 + H)^2}-\rho^2\right)\cos{\rho H}+\frac{4G_0 \rho}{(1 + H)^3}\sin{\rho H}\right]\left(H_{XX} + H_{YY}\right)
 \label{QSEterms1} \\ 
&\equiv& q_{31}\left(H;\; G_0,G_1,\rho\right)\left(H_{XX} + H_{YY}\right); \nonumber
\end{eqnarray}
this is another $\gamma^{(QSE)}$ contribution, which originates in $\bar \nabla_s^2 \bar \mu_{wet}$. 
\begin{equation}
f_5= -\epsilon_\gamma\left[a_{00} \left(H_{XXXX} + 2H_{XXYY} +H_{YYYY}\right)+2 a_{20} \left(H_{XXXX} + H_{XXYY}\right)  
 +2 a_{02} \left(H_{YYYY}+H_{XXYY}\right)\right];
\label{f_50}
\end{equation}
this is $\gamma^{(Anis)}$ contribution, where the terms that are proportional to $a_{00}$ originate in $\bar \nabla_s^2 \bar \mu_{\kappa}$, and the terms that are proportional to
$a_{02}$ and $a_{20}$ originate
in $\bar \nabla_s^2 \bar \mu_{anis}$.  Notice that the anisotropy coefficients $a_{12},\ a_{22},\ a_{30},\ a_{40},\ a_{04}$ do not enter the PDE. 
\begin{equation}
f_6=\Delta\left(H_{XXXXXX} + 3 H_{XXXXYY} + 3 H_{XXYYYY} + H_{YYYYYY}\right);
\label{f_5}
\end{equation}
this is another $\gamma^{(Anis)}$ contribution (the regularization), which originates in $\bar \nabla_s^2 \bar \mu_{h.o.t.}$. We also introduced 
the coefficient functions $q_{11}(H; \ldots)\equiv -\gamma^{(SS)}(H; \ldots),\ q_{12}(H; \ldots),\ q_{21}(H; \ldots)\equiv \gamma^{(QSE)}(H; \ldots),\ q_{31}(H; \ldots)$ 
in order to simplify the appearance of equations in the next section; here the ellipsis again denotes the dimensionless parameters.

\section{Linear Stability Analysis}

\label{LSA}

In the vein of a standard linear stability analysis, the planar surface $H=H_0$ of the as-deposited film is perturbed by a 
small perturbation $\xi(X,Y,T)$, i.e. $H(X,Y,T)=H_0+\xi(X,Y,T)$. Substitution of this form into Eq. (\ref{PDEcompact}) and linearization in $\xi$ yields the coefficient functions $q_{ij}(H_0; \ldots)$, allowing to study the effects of a film thickness on film stability. Also, the derivatives of $H$ are replaced by the corresponding derivatives of $\xi$. Then $\xi$ is taken in the form of a
wave-like perturbation $\xi=b\ e^{\omega T}e^{i\left(k_1X+k_2Y\right)}$, where $k_1$ and $k_2$ are wave numbers along the $X$ and $Y$ directions, $\omega$ the growth (or decay) rate, 
and $b\ll 1$. This finally results in the following expression for $\omega$ (the dispersion relation):
\begin{eqnarray}
\omega\left(k_1,k_2\right) &=& \left(q_{11}- q_{21}-a_{00}\epsilon_\gamma-1\right) \left(k_1^4 + 2k_1^2k_2^2 + k_2^4\right) -2\epsilon_\gamma \left[a_{20}\left(k_1^4 + k_1^2k_2^2\right) 
+a_{02}\left(k_2^4 + k_1^2k_2^2\right)\right] \label{OMEGA} \\
&-& \left(q_{12}+q_{31}\right) \left(k_1^2 + k_2^2\right) -\Delta\left(k_1^6 + 3k_1^4k_2^2 + 3k_1^2k_2^4 + k_2^6\right) \nonumber.
\end{eqnarray}
Here we omitted the arguments of $q_{ij}$'s. 

Next, we proceed to analyze three important situations: QSE only; QSE and the anisotropy; QSE, the anisotropy and the surface stress.

\subsection{QSE}
\label{LSA_QSE}

\begin{figure}[h]
\centering
\includegraphics[width=6.0in]{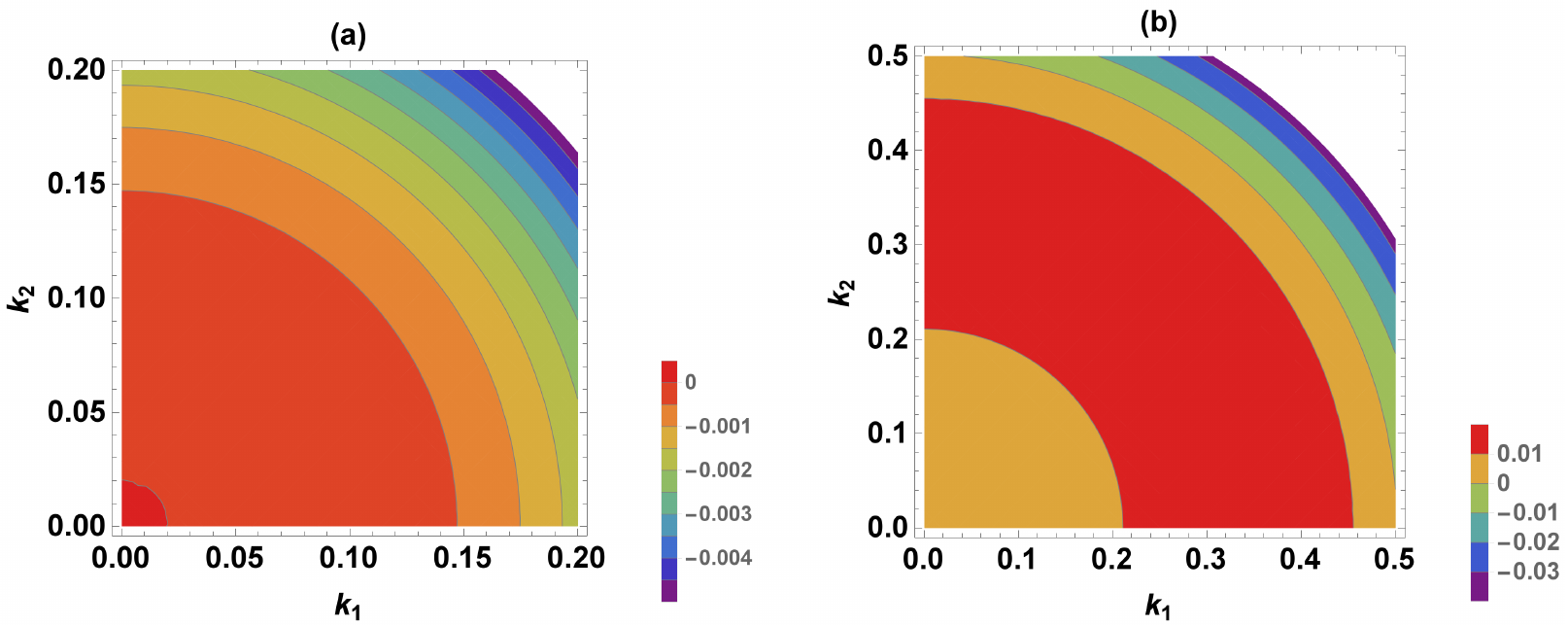}
\caption{(Color online.) Contour plots of the perturbation growth rate $\omega\left(k_1,k_2\right)$ for the QSE case, Eq. (\ref{omega_qse}),  where $R_{\gamma_0}=G_0=G_1=0.1,\ H_0=5$. 
(a) $\rho=0$; (b) $\rho=10$. 
}
\label{FigOmega_Iso_NoStress}
\end{figure}

When only QSE is operative, the dispersion relation (\ref{OMEGA}) takes the form:
\begin{equation}
\omega\left(k_1,k_2\right) = -\left(q_{21}+1\right) \left(k_1^4 + 2k_1^2k_2^2 + k_2^4\right) -q_{31} \left(k_1^2 + k_2^2\right).
\label{omega_qse}
\end{equation}
Notice that the Mullins' contribution $-\left(k_1^4 + 2k_1^2k_2^2 + k_2^4\right)$ always must be retained in $\omega$, since the surface of a film is curved.
One can now see that the surface is stable, if $q_{21}+1,\ q_{31} \ge 0$, and it is unstable with respect to growth of a long-wave perturbations, i.e. those with wave numbers
$k_1 < k_{1c},\ k_2<k_{2c}$, if  $q_{31} < 0$ and  $q_{21}+1 > 0$. The latter two conditions ensure that small (large) wave numbers grow 
(decay), thus the cut-off values $k_{1c}$ and $k_{2c}$ exist. A typical growth rate at a fixed unstable thickness is plotted in Figure \ref{FigOmega_Iso_NoStress}. 
It can be noticed by comparing the panels (a) and (b) of this Figure that the QSE oscillation makes the film more unstable, since the interval of the unstable 
wave numbers is wider when $\rho>0$ and the maximum growth rate $\omega_{max} = \omega\left(k_{1max},k_{2max}\right)$ is one order of magnitude larger. Since the (equal) wave numbers 
$k_{1max}$ and $k_{2max}$ are larger than in the $\rho=0$ case, the fastest growing perturbation has shorter wavelength.

From the stability conditions $q_{21}+1,\ q_{31} \ge 0$ we obtain, at $\rho=0$:
\begin{eqnarray}
G_1 &\le& \frac{G_0}{1 + H_0} + 1 + H_0, 
\label{c2} \\
G_1 &\le& \frac{3G_0}{1+H_0}. 
\label{c1}
\end{eqnarray}
The second of these conditions is (\ref{StabCondThermodyn}) in Sec. \ref{QSE_sect}, which was obtained from the thermodynamic argument. 
Notice that since $0 < G_0,\ G_1 < 1$, the first condition holds for any
$H_0$. The film is unstable when (\ref{c1}) does not hold. 

\begin{figure}[h]
\centering
\includegraphics[width=3.0in]{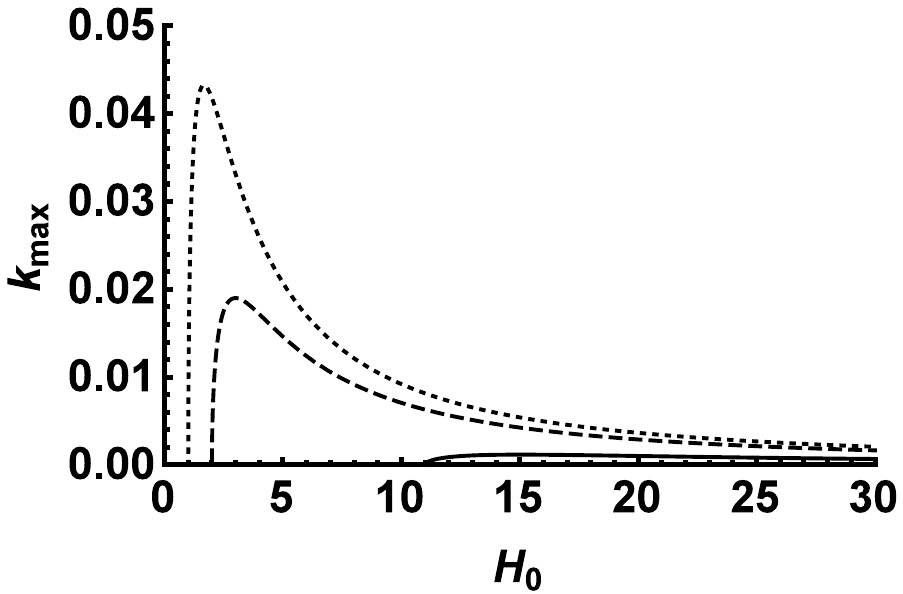}
\caption{Long-wave instability for the QSE case;  $\rho=0,\ R_{\gamma_0}=G_0=0.1$. Curves show $k_{max}$ vs. $H_0$ for  
$G_1=0.025,\ 0.1,\ 0.15$  (solid, dashed, and dotted lines, respectively). 
}
\label{kmax1D_isotropic_nostress_A=0}
\end{figure}

\begin{figure}[h]
\centering
\includegraphics[width=6.0in]{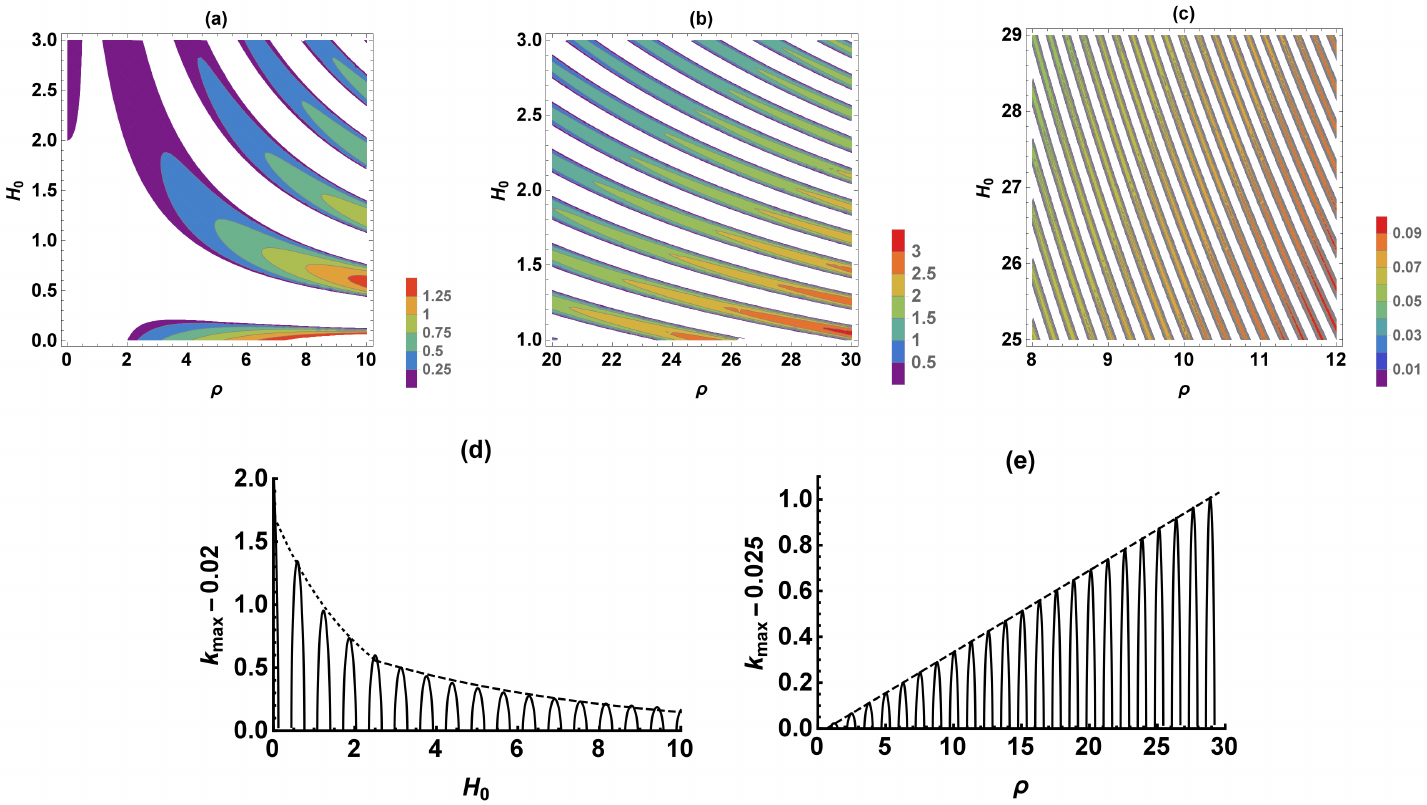}
\caption{(Color online.) Long-wave instability for the QSE case;  $R_{\gamma_0}=G_0=G_1=0.1$. 
Panels (a)-(c) show the zooms into different regions of the contour plot of $k_{max}$; the film is stable in the white regions, i.e. $\omega\left(k_1,k_2\right) \le 0$ for all non-negative wave numbers. 
Panels (d) and (e) show the cross-section of the contour plot in the panel (a) by the vertical line $\rho=10$, and by the horizontal line $H_0=5$, respectively 
(the curves are slightly shifted down for better view). See the text for the descriptions of the dotted and dashed curves.
}
\label{Kmax_vs_psi_rho_isotropic_nostress_A=10}
\end{figure}

If instead of the described two-dimensional (2D) problem setting one chooses a simpler one-dimensional (1D) setting, then $k_2=0$ in Eq. (\ref{omega_qse}) and the growth rate $\omega_{max}$ and the wave number $k_{1max}\equiv k_{max}$  of the fastest growing perturbation 
can be easily determined analytically from this equation. $k_{max}$ also can be detected in experiment, since this
mode will dominate shortly after the evolution starts.
In the 2D setting $k_{1max}$, $k_{2max}$ and  $\omega_{max}$ may be computed.
However, due to a symmetry of Eq. (\ref{omega_qse}) with respect to $k_1$ and $k_2$ in the isotropic case, 
the variations of $k_{1max}$, $k_{2max}$, and the maximum growth rate with respect to the parameters are the same as the corresponding variations of a ``1D" $k_{max}$ and 
$\omega_{max}$. (From here on, and to the paper's end, $k_{max}$ will mean a ``1D" $k_{max}$.) Also in this case $k_{1max}=k_{2max}=k_{max}$.  All just said is true in some anisotropic situations, such as for example the [111] and [001] surface orientations; here, since $a_{02}=a_{20}$ (see the Table 1 and Eq. (\ref{f_50})) the symmetry is unbroken.
Thus with a goal of better understanding key dependencies on the parameters, where possible we will use a simpler 1D setting and calculate and then plot $k_{max}$ or $\omega(k_1,0)$
(see Figures \ref{kmax1D_isotropic_nostress_A=0} - \ref{kmax_vs_eps_gamma_H0_qse_and_anisotropy111}, \ref{kmax_vs_epsstar_H0_qse_anisotropy_stress_111} - \ref{omega1D_tensile_H0=2_vs_A}). 
In reference to these Figures, the ``wave length" of the fastest growing 2D perturbation $\lambda_{max}=2\pi/\sqrt{k_{1max}^2+k_{2max}^2}
=\pi \sqrt{2}/k_{max}$.

Therefore, using the 1D setting we obtain 
\begin{equation}
\omega_{max} = -\left(q_{21}+1\right)k_{max}^4 -q_{31}k_{max}^2,\quad k_{max} = \sqrt{-q_{31}/2(1+q_{21})},\quad k_c=\sqrt{2}k_{max}.
\end{equation}
For $\rho=0$ case, Figure \ref{kmax1D_isotropic_nostress_A=0} shows $k_{max}$ vs. $H_0$ for various  
ratios $G_1/G_0$.  The film is stable at $H_0<11$ for $G_1=0.025$, at
$H_0<2$ for $G_1=0.1$, and at $H_0<1$ for $G_1=0.15$ (since $k_{max}$ vanishes); this means that the decrease of the component of $\gamma^{(QSE)}$ that is due to 
charge spilling results in thicker stable films. This is also evident from the stability condition (\ref{StabCondThermodyn}) (or (\ref{c1})). \footnote{ In section 5 of his paper, Hirayama \cite{Hirayama} cites the experiments where the shift of the critical thickness is arguably achieved 
through enhancement of charge spillage at the interface after the layer of Al is inserted between Ag film and Si substrate.}   For all admissible $G_1$ values (such that $0< G_1 < 2G_0$) the film is unstable at large 
$H_0$ (notice that $k_{max}$ is finite and small). 

When $\rho > 0$ (the QSE oscillation), the analytical conditions similar to (\ref{c1}) and  (\ref{c2}) are impossible to derive. 
Notice that stability in this situation depends not only on the product $\rho H_0$, but also on $H_0$ and $\rho$ separately, see Eqs. 
(\ref{QSEterms}), (\ref{QSEterms1}).
In the panels (a)-(c) of Figure \ref{Kmax_vs_psi_rho_isotropic_nostress_A=10} the contour plot of the 
function $k_{max}(\rho,H_0)$ is shown. First, one notices that for any fixed $\rho>2$, as $H_0$ increases, the film is alternatingly unstable or stable. 
At larger $\rho$ the unstable bands become more flat, i.e. their slopes decrease (panel (b)),
and at larger $H_0$ more unstable bands appear in any fixed interval of $\rho$, i.e. their density increases; also they are more vertical (larger slopes, see the panel (c)). Thus we expect that at $\rho\gg 1$ the bands are horizontal and separated, and at $H_0\gg 1$, there is a longwave 
instability at any $\rho$ with $k_{max}\approx const.$ (notice the uniform coloring in the panel (c)). This is expected, since in the limit $H\rightarrow \infty$:
$\gamma^{(QSE)}\approx R_{\gamma_0}-\frac{G_1}{1+H}$, i.e. the QSE oscillation is extinct, 
and thus $\partial^2 \gamma^{(QSE)}/\partial H^2 <0$ for all positive $G_1$ values (the instability by the thermodynamic criterion). 
On the other hand, when $\rho\gg 1$ and $H$ is fixed and not very large, 
the dominant term in $\partial^2 \gamma^{(QSE)}/\partial H^2$ is $-G_0 \rho^2 \cos{\rho H}/\left(1+H\right)^2$. Thus in this limit the film is unstable whenever
$\cos{\rho H}>0$, which at fixed $\rho$ and variable $H$ produces alternating stable and unstable horizontal bands.
To this end, comparing the panel (d) with Figure \ref{FigGammaQSE} (which is plotted also at $\rho=10$), it can be seen that the positions along the $H_0$-axis of the stable (unstable) bands 
closely match the $H$-coordinates of the points of minimum (maximum) of the $\gamma^{(QSE)}(H)$ curve. The envelope of the function $k_{max}\left(H_0\right)$ in this panel
is closely fitted, as shown, by two rather disparate exponential functions of the form $a\exp{(-ct)}$; for small $H_0$ values, $a=1.737$, $c=0.438$; and for large $H_0$ values, $a=0.872$, $c=0.165$. In Figure \ref{Kmax_vs_psi_rho_isotropic_nostress_A=10}(e) the envelope is the linear function $k_{max}=0.036 \rho + 0.0016$.

To wrap up the discussions in this section, we remark that the alternation of stable and unstable film thicknesses, or a more complicated beating patterns, have been predicted by the ``electronic growth" theory and 
seen in experiment 
\cite{Smith,Zhang,Hirayama,Ozer,Han1}.
We also remark that the surface deformation that emerges from the long-wave instability is bi-continuous, $H\sim e^{\omega_{max}t}\cos{k_{1max}X_1}\cos{k_{2max}X_2}$, since $k_{1max}$ and  $k_{2max}$ are non-zero (and the same value, as we discussed above);
see Figure \ref{FigOmega_Iso_NoStress}.


\subsection{QSE and anisotropy}
\label{LSA_QSE_anis}

Some level of the surface energy anisotropy is always present in a crystal film system. Pinholes that form in the ultra-thin Ag films  \cite{Sanders} and in thicker (80 nm) Ni films
\cite{Rabkin} appear faceted and, though uncommon, metal islands also may grow into a highly faceted shapes \cite{Wall,Bennett}.
In this section we analyze how the combination of QSE and anisotropy affects the film stability.
When the anisotropy is accounted for, the contributions $f_5$ and $f_6$ are present at the right-hand side of Eq. (\ref{PDEcompact}), unless the anisotropy is mild, i.e. the anisotropy strength parameter $-1/(a_{00}+a_{20}+a_{02}) < \epsilon_\gamma < 0$.
In the latter case
the regularization parameter $\Delta=0$ and the $f_6$-term vanishes. 
When the anisotropy is strong, $\epsilon_\gamma \le -1/(a_{00}+a_{20}+a_{02})\equiv \epsilon_{\gamma c}$, a faceting instability is expected to emerge;
here $\epsilon_{\gamma c}$ is the critical value.  A faceting instability was theoretically studied in several 
papers for different situations, which include film growth, substrate wetting, and heteroepitaxial stress, see for instance Refs. \cite{LiuMetiu,GDN,SGDNV,GolovinPRB2004,LevinePRB2007,Korzec,Korzec1,K,K1,K2,LLRV,KTL}.

We only consider the QSE oscillation case, $\rho>0$, and vary only $\epsilon_\gamma$ and 
$H_0$. Value of the latter parameter depends on the amount of a film material deposited on a substrate, and the former parameter depends on the temperature and the material itself.
In addition, we will compare the [111] and [011] surface orientations; for the latter orientation the coefficients $a_{20}$ and $a_{02}$ in Eq. (\ref{f_50}) differ in the sign and magnitude (a broken symmetry), see the Table 1. This makes the stability character quite different from the [111] case.

\begin{figure}[h]
\centering
\includegraphics[width=6.0in]{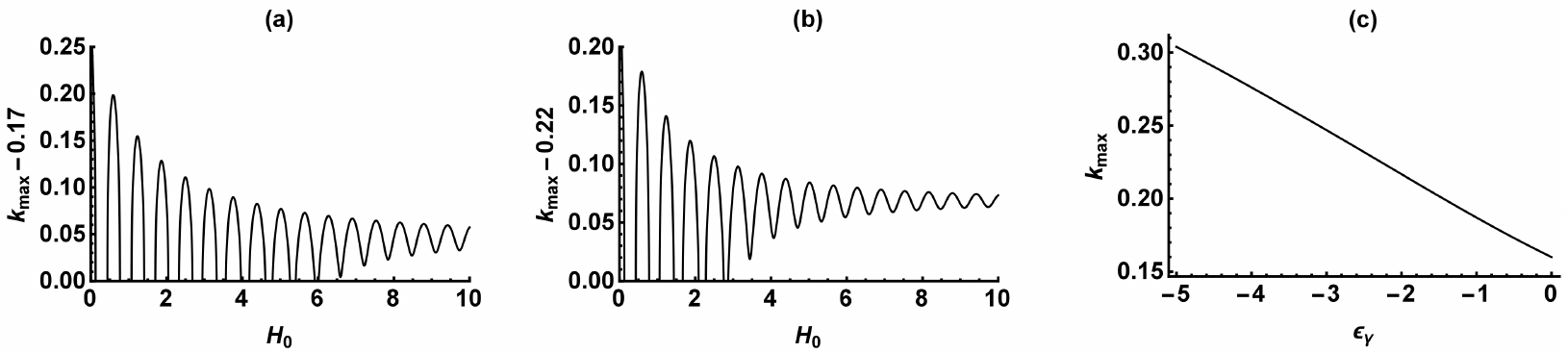}
\caption{Long-wave instability for QSE$+$anisotropy case and [111] surface orientation. $\rho=10,\ R_{\gamma_0}= G_0=G_1=0.1$, and
$\Delta=83$.
(a) $\epsilon_\gamma=-3 < \epsilon_{\gamma c}=-1/3$; (b) $\epsilon_\gamma=-5$; (c) $H_0=5$.
}
\label{kmax_vs_eps_gamma_H0_qse_and_anisotropy111}
\end{figure}

For the [111] surface $k_{1max}=k_{2max}=k_{max}$ due to the symmetry of the anisotropy coefficients, and in Figure \ref{kmax_vs_eps_gamma_H0_qse_and_anisotropy111}(a,b)
we plot $k_{max}$ vs. $H_0$ for $\epsilon_\gamma$ and other parameters fixed. One can see how increasing the anisotropy (i.e. $|\epsilon_\gamma|$ becomes larger) 
promotes instability by eliminating $H_0$ values where the surface would 
be stable ($k_{max}=0$) if there was no anisotropy or if the anisotropy was weak. 
In the panel (c) of this Figure $k_{max}$ is plotted vs. $\epsilon_\gamma$ for fixed $H_0$ and other parameters. It can be seen that
increasing $|\epsilon_\gamma|$ slowly shifts the instability (in the linear fashion) toward shorter wavelengths. Overall, $k_{max}$ is only weakly sensitive to the modest
changes of the anisotropy strength.

\begin{figure}[h]
\centering
\includegraphics[width=6.0in]{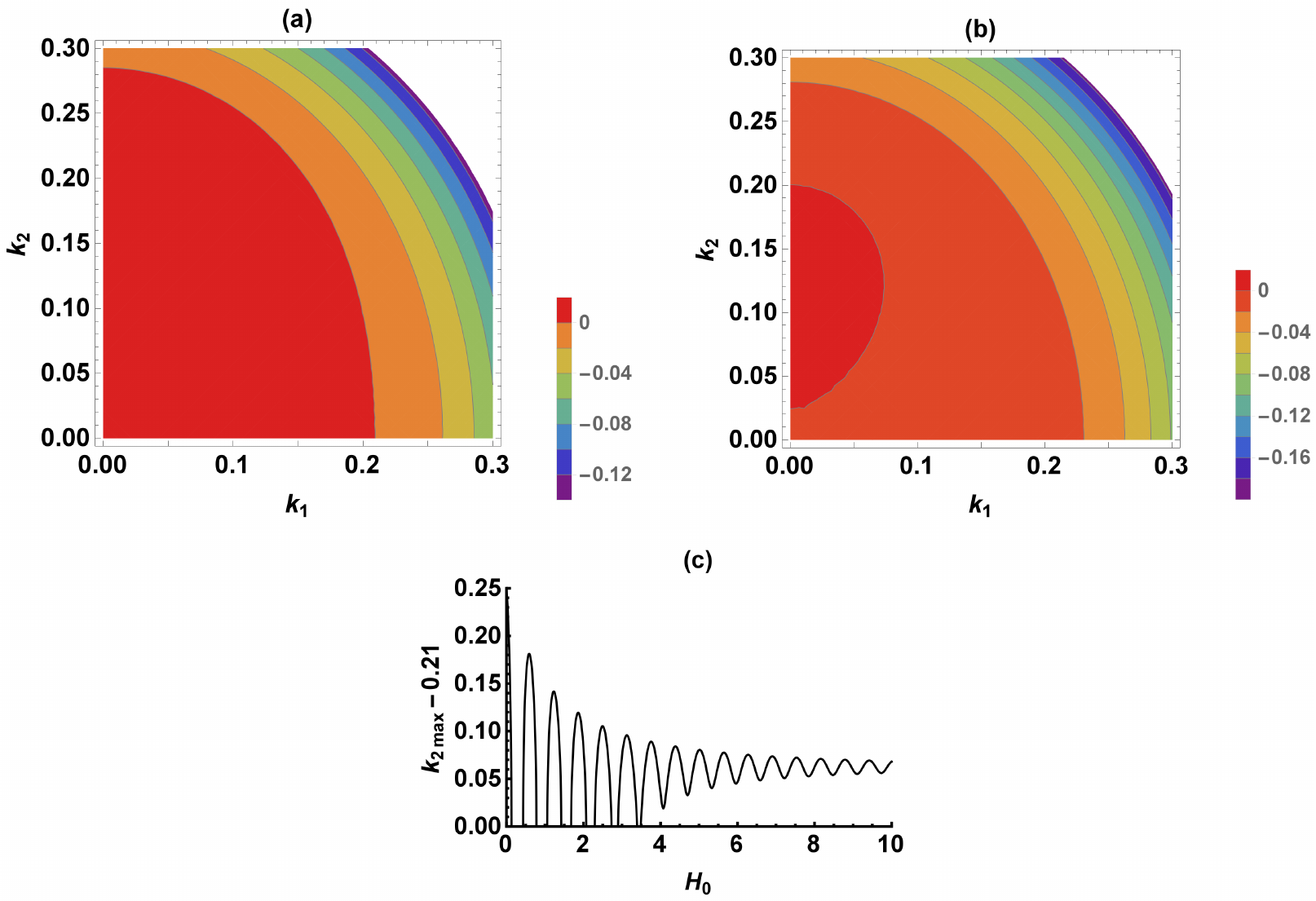}
\caption{(Color online.) (a,b): Contour plots of the perturbation growth rate $\omega\left(k_1,k_2\right)$ for the QSE$+$anisotropy case and the [011] surface orientation, where 
$\rho=10,\ R_{\gamma_0}=G_0=G_1=0.1$, $\epsilon_\gamma=-1 < \epsilon_{\gamma c} = -2/3$ (strong anisotropy), and
$\Delta=83$. (a) $H_0=5$; (b) $H_0=35$. (c): The wave number $k_{2max}$ 
(at which $\omega$ attains a maximum) vs. $H_0$; here $\epsilon_\gamma=-3$.
}
\label{kmax_vs_eps_gamma_H0_qse_and_anisotropy011}
\end{figure}

For the [011] surface, the panels (a) and (b) of Figure \ref{kmax_vs_eps_gamma_H0_qse_and_anisotropy011} show 
how  
$k_1$ values are steadily eliminated from the spectrum of the unstable wave numbers with the increase of $H_0$. 
At weak anisotropy ($-2/3<\epsilon_\gamma<0,\ \Delta=0$) the analysis of the conditions for the existence of the local maximum, $\partial \omega\left(k_1,k_2\right)/\partial k_1=0,\ 
\partial \omega\left(k_1,k_2\right)/\partial k_2=0$ (see Eq. (\ref{OMEGA})) is possible, since these equations are quadratic in $k_1$ and $k_2$, respectively. 
It shows that for $a_{00}=1/2$, $a_{20}=-1$ and $a_{02}=2$ (and other characteristic parameters) there is no real-valued solution $k_1$ (other than $k_1=0$).
Also, if one chooses any initial condition $(k_1,k_2)$ inside the squares (Figure \ref{kmax_vs_eps_gamma_H0_qse_and_anisotropy011}(a) or (b)), then Mathematica's \cite{Mathematica} numerical
maximization function \emph{FindMaximum} converges to a solution $(0,k_{2max})$, where $k_{2max}$ is a numerical value.
This means that the fastest growing surface deformation is one-dimensional, i.e. $H\sim e^{\omega_{max}t}\cos{k_{2max}X_2}$, which describes a long rolls with their axes 
(spaced $2\pi/k_{2max}$ units apart) directed along $X_1$-axis.
The panel (c) of Figure \ref{kmax_vs_eps_gamma_H0_qse_and_anisotropy011} is similar to Figure \ref{kmax_vs_eps_gamma_H0_qse_and_anisotropy111}(a); comparing to the latter figure, one notices that in the [011] case a smaller supercritical anisotropy would be sufficient for 
instability in the entire plotted interval of $H_0$. 

\subsection{QSE, anisotropy, and surface stress}
\label{LSA_QSE_anis_stress}

In this section we briefly discuss the film stability in the presence of all three major physical effects: QSE,  the anisotropy and the surface stress. The number of the dimensionless parameters is large in this case, thus we only aim at revealing some most prominent features of the surface instability. The discussion of the 
[111] surface suffices in this regard.

\begin{figure}[h]
\centering
\includegraphics[width=6.0in]{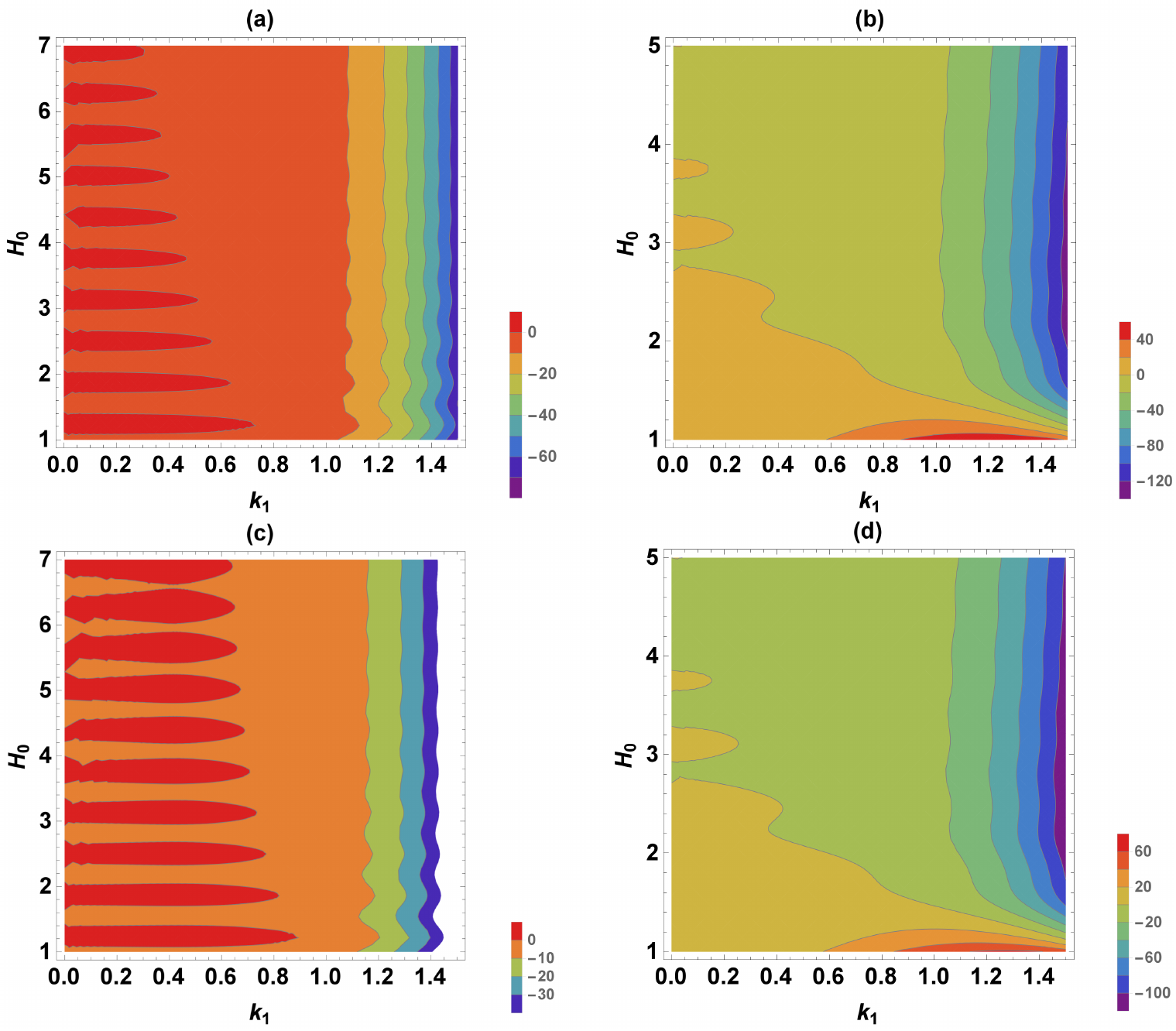}
\caption{(Color online.) Contour plots of $\omega\left(k_1,0\right)$ for the QSE+anisotropy+surface stress case, where the surface orientation is [111], the stress is compressive ($c_s=-1$), 
$\rho=10,\ R_{\gamma_0}=G_0=G_1=0.1$, and other parameters as in the caption to Figure \ref{FigGammaSS} (except $\Gamma P=9.6\times 10^{-6}$ erg). 
(a) $\epsilon_*=10^{-5}$, $\epsilon_\gamma=-0.1$. (b) $\epsilon_*=10^{-3}$, $\epsilon_\gamma=-0.1$. 
Parameters in (c), (d) are the same as in (a), (b), respectively, except $\epsilon_\gamma=-1$ (strong anisotropy). In the white areas the film is stable.
}
\label{kmax_vs_epsstar_H0_qse_anisotropy_stress_111}
\end{figure}

\begin{figure}[h]
\centering
\includegraphics[width=6.0in]{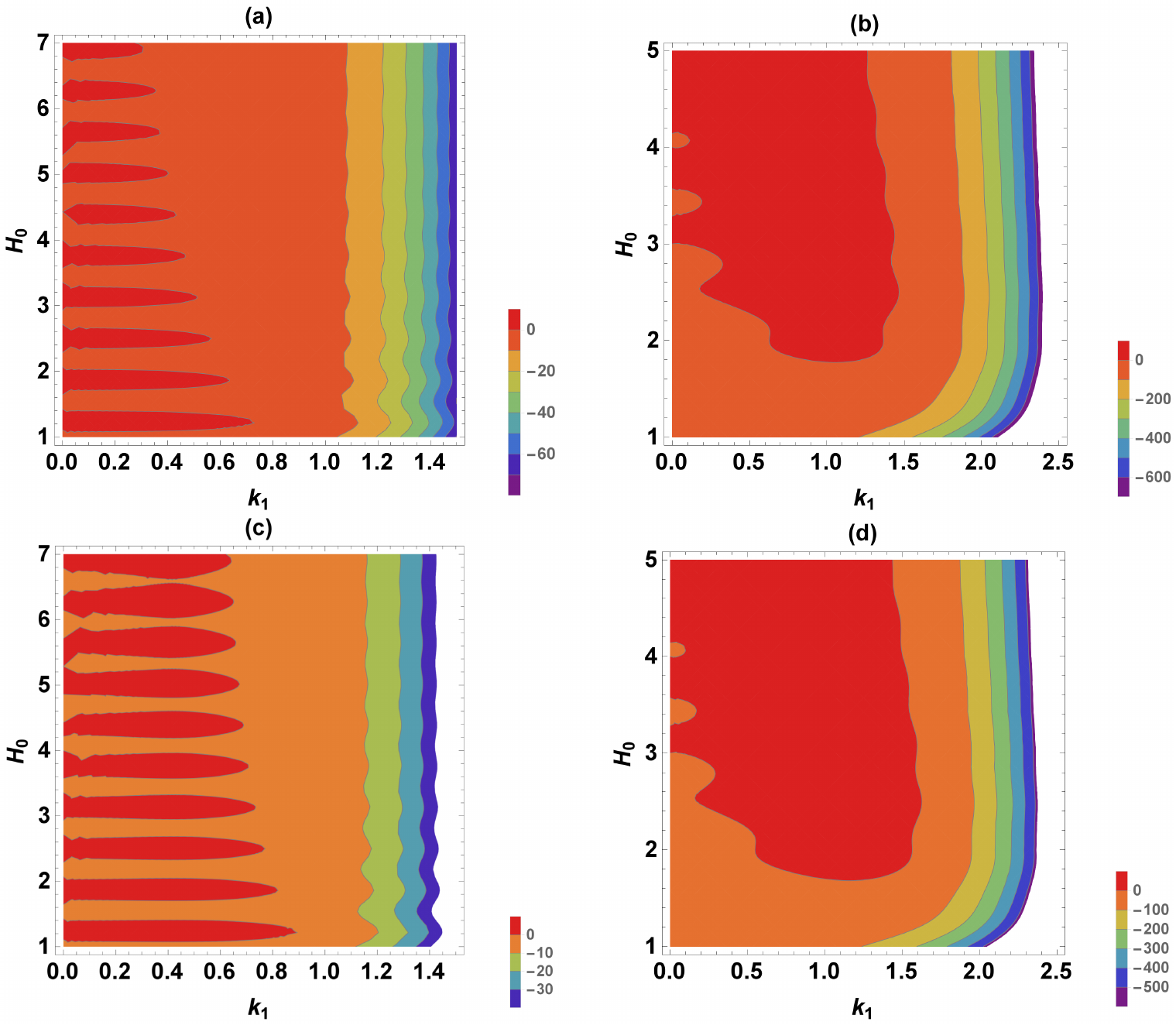}
\caption{(Color online.) 
Same as Figure \ref{kmax_vs_epsstar_H0_qse_anisotropy_stress_111}, except that the surface stress is tensile ($c_s=1$).
Notice that the panels (a) and (c) coincide with the panels (a), (c) in Figure \ref{kmax_vs_epsstar_H0_qse_anisotropy_stress_111}. 
In the white areas the film is stable.}
\label{kmax_vs_epsstar_H0_qse_anisotropy_stress_111_tensile}
\end{figure}

\begin{figure}[h]
\centering
\includegraphics[width=2.5in]{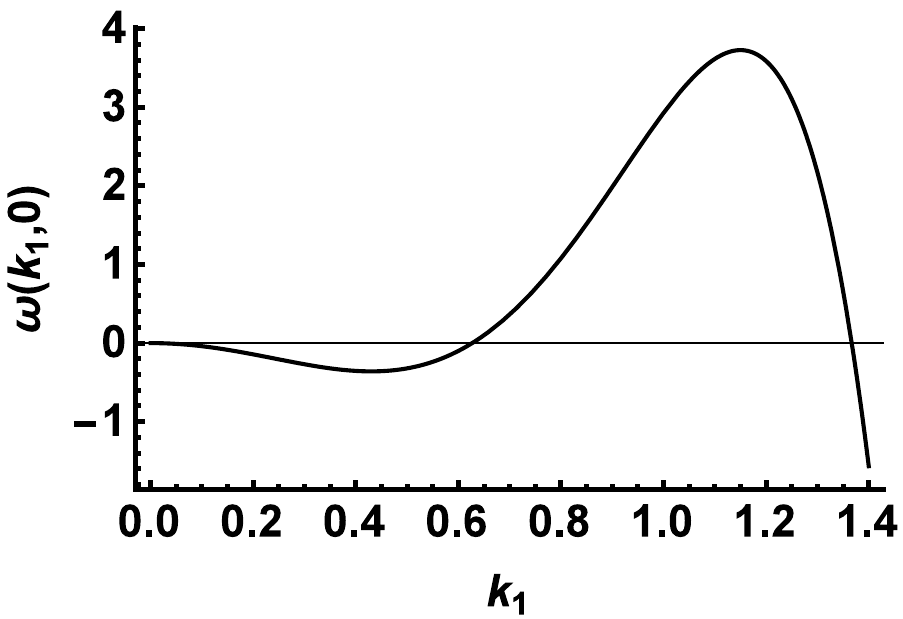}
\caption{Cross-section of the contour plot in Figure \ref{kmax_vs_epsstar_H0_qse_anisotropy_stress_111_tensile}(b) by the line $H_0=2$. }
\label{slice_of_omega1D}
\end{figure}

\begin{figure}[h]
\centering
\includegraphics[width=7.0in]{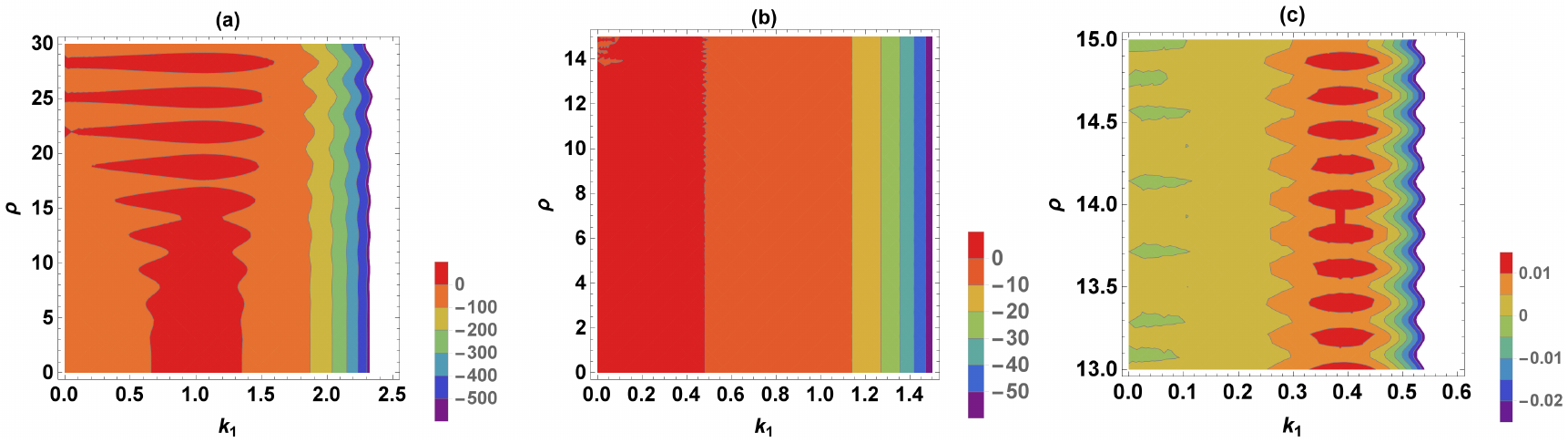}
\caption{(Color online.) 
Panels (a) and (b) are the same as Figure \ref{kmax_vs_epsstar_H0_qse_anisotropy_stress_111_tensile}(b), except here $\rho$ varies and $H_0=2$ in (a), $H_0=30$ in (b).
Panel (c) shows the zoom into the upper-left corner of the panel (b).}
\label{omega1D_tensile_H0=2_vs_A}
\end{figure}

We use the 1D approach since the anisotropy of the [111] surface does not break the symmetry even in the presence of the surface stress.
The plots of $\omega\left(k_1,0\right)$ are shown in Figures \ref{kmax_vs_epsstar_H0_qse_anisotropy_stress_111} - \ref{omega1D_tensile_H0=2_vs_A}. 

Comparing Figures \ref{kmax_vs_epsstar_H0_qse_anisotropy_stress_111} and \ref{kmax_vs_epsstar_H0_qse_anisotropy_stress_111_tensile} we notice first that at the small value of the stress parameter $\epsilon_*$ the anisotropy dominates irrespective of whether the stress is compressive
or tensile (Figure \ref{kmax_vs_epsstar_H0_qse_anisotropy_stress_111}(a,c) and Figure \ref{kmax_vs_epsstar_H0_qse_anisotropy_stress_111_tensile}(a,c)). 
Unstable ``fingers" appear when the film thickness is roughly a value at which $\gamma^{(QSE)}$ attains a maximum (Figure  \ref{FigGammaQSE}), and these fingers are
longer at larger anisotropy. With the increase of the film thickness the fingers become shorter and more narrow at weak anisotropy and eventually they disappear, i.e. at large
$H_0$ the stabilizing impact of the compressive stress overpowers the destabilizing impact of QSE and weak anisotropy. At strong anisotropy the fingers are longer and wider, and they merge when the thickness is sufficiently large, i.e the destabilization by strong anisotropy and QSE overpowers the stabilization by the compressive stress.  Thus at weak anisotropy the large thicknesses are stable, but at strong anisotropy they are unstable. 
Increasing the compressive stress at either a weak or a strong anisotropy (Figures \ref{kmax_vs_epsstar_H0_qse_anisotropy_stress_111}(b,d)) has the effect of nearly equally stabilizing the large thicknesses; notice that  the only apparent difference between the panels (b) and (d) is the magnitude of $\omega$ (it is larger at strong anisotropy, 
panel (d)). Thus large compressive stress prevails over the anisotropy and QSE. The instability in Figure \ref{kmax_vs_epsstar_H0_qse_anisotropy_stress_111} has a long-wave character.

Quite interestingly, increasing the tensile stress changes the character of the instability from the long-wave 
(Figures \ref{kmax_vs_epsstar_H0_qse_anisotropy_stress_111_tensile}(a,c)) to the short-wave (Figures \ref{kmax_vs_epsstar_H0_qse_anisotropy_stress_111_tensile}(b,d)). 
The plot of the growth rate at $H_0=2$ is shown in Figure \ref{slice_of_omega1D}; all such plots at any $H_0<3$ and at $H_0\approx 3.2, 4.05$ are qualitatively similar.
Notice that $\omega\left(k_1,0\right) < 0$ at the large \textit{and} small 
wave numbers.
The short-wave instability may change the nonlinear evolution of the film during dewetting from coarsening to the formation of spatially regular patterns \cite{GolovinPRB2004}. 
In Figure \ref{omega1D_tensile_H0=2_vs_A}(a) it can be seen that at $H_0=2$ the short-wave instability manifests for all $\rho < \sim 17$, and then at 
$\rho$ around 19, 22, 25, 28, etc. When $H_0>\sim 4$ the instability is long-wave for $\rho$ below some threshold value ($\approx 14$ at $H_0=30$, see Figures \ref{omega1D_tensile_H0=2_vs_A}(b,c)), and short-wave for larger $\rho$ values. Also one can notice by comparing
Figures \ref{kmax_vs_epsstar_H0_qse_anisotropy_stress_111_tensile}(b,d) that strong anisotropy slightly widens the domain of unstable wavenumbers.


\section{Conclusions}

\label{Concl}

We studied the linear stability of a surface of a supported ultra-thin metal film using a PDE-based model that incorporates
quantum size effect, the surface energy anisotropy, and the surface stress. 
By a systematic comparison, a contribution of each effect into the overall stability (or instability) can be discerned. The following major trends 
are revealed (some for the first time): 

\begin{enumerate}

\item For QSE alone, without the QSE oscillation (the dimensionless oscillation wavenumber $\rho=0$, see Eq. (\ref{gammaQSE2})): a film is unstable when its dimensionless thickness  
$H_0 > 3G_0/G_1-1=H_{0c}$. Here $G_0,\; G_1$ are the parameters that define  
the ``intensity" of the electrons quantum confinement in the film and the charge spilling at the film/substrate interface, respectively.
This is the usual ``thermodynamic" condition for instability. Weakening of the charge spilling by reducing $G_1$ results in thicker stable films.

\item For QSE alone, with the QSE oscillation $(\rho> 0)$:
For moderate and large $\rho$ values ($\rho>2$ is typical; notice that $\rho\sim 10$ corresponds to most metals that were studied in the experiments), as $H_0$ increases from zero there emerge the alternating stable and unstable intervals (bands) of film 
thicknesses that follow the
$H$-coordinates of the
minimum and maximum points on the $\gamma^{(QSE)}(H)$ curve. At $\rho\gg 1$ the bands are horizontal and separated, and at $H_0\gg 1$, there is a longwave 
instability at any $\rho$ with $k_{max}\approx const.$ If $H_0$ and $\rho$ are such that the film is unstable, then the instability wave length increases as $H_0$ increases and $\rho$ decreases.
If a film is unstable at some $H_0$ with or without the QSE oscillation, then the spectrum of the unstable wave numbers and the maximum perturbation growth rate are larger in the former case.

\item For QSE (with the QSE oscillation) and anisotropy: As the oscillation imposes the alternation of  stable and unstable thickness bands, 
increasing the anisotropy contributes to instability by eliminating stable bands (starting with large thickness bands).
Also, if a film of a certain thickness $H_0$ is unstable without the anisotropy and the QSE oscillation, then ``turning on" the anisotropy makes the most dangerous wavelength smaller. 
The [011] surface is more
unstable than the [111] surface. At least for our choice of the most typical set of parameters, the deformation of the [011] surface has the form of the parallel rolls (the 1D mode), while the deformation of the [111] surface is bicontinuous (the 2D mode).

\item For QSE with the QSE oscillation, the anisotropy, and the surface stress: 
We found different instability characters depending on whether the stress is compressive (the long-wave instability for all film thicknesses) or tensile 
(a combination of the short-wave and long-wave instabilities).
Increasing the compressive stress eliminates large unstable film thicknesses, i.e. thick  films are stabilized  by 
the compressive stress.  On the contrary, thin films are destabilized by the compressive stress.
The tensile stress has the opposite effect. 
As such, it reinforces the actions of
QSE and anisotropy and as the result, the short-wave instability emerges at small film thicknesses 
when $\rho$ is small to moderate, and at large thicknesses when $\rho$ is large.

\end{enumerate}

We hope that these results can be used to better understand the dewetting mechanisms of an ultra-thin metal films. 

\bigskip
\noindent
\textit{Acknowledgments.}$\;$ The author thanks Olivier Pierre-Louis (University of Lyon) for making available his unpublished recent derivation of the electronic contribution to
the surface energy.  The anonymous Referee is acknowledged for the expert and very thorough multi-stage review of the paper's manuscript.

\end{document}